\documentclass[journal,10pt,twocolumn,twoside]{IEEEtran}

\usepackage{graphicx}
\usepackage{amsmath, amsthm, amssymb, amsfonts, cancel}
\usepackage{algorithm, algorithmic, ifsym}

\usepackage{subfig}

\usepackage{float}

\newcommand\blfootnote[1]{%
  \begingroup
  \renewcommand\thefootnote{}\footnote{#1}%
  \addtocounter{footnote}{-1}%
  \endgroup
}

\def\Cov{{\rm Cov}}

\def\bn{ \textbf{n} }

\def\bA{ {\mathbf{A}} }
\def\bB{ {\mathbf{B}} }
\def\bC{ {\mathbf{C}} }
\def\bD{ {\mathbf{D}} }
\def\bE{ {\mathbf{E}} }
\def\bF{ {\mathbf{F}} }
\def\bG{ {\mathbf{G}} }
\def\bH{ {\mathbf{H}} }
\def\bI{ {\mathbf{I}} }

\def\bK{ {\mathbf{K}} }

\def\bM{ {\mathbf{M}} }

\def\bP{ {\mathbf{P}} }

\def\bS{ {\mathbf{S}} }

\def\bU{ {\mathbf{U}} }
\def\bV{ {\mathbf{V}} }
\def\bW{ {\mathbf{W}} }

\def\bZ{ {\mathbf{Z}} }

\def\bSigma{ {\mathbf{\Sigma}} }
\def\bDelta{ {\mathbf{\Delta}} }
\def\bLambda{ {\mathbf{\Lambda}} }
\def\bGamma{ {\mathbf{\Gamma}} }

\def\nn{{ \parallel   }}
\def\RR{{ \mathbb{R}  }}
\def\CC{{ \mathbb{C}  }}

\def\EE{{ \mathbb{E}  }}

\def\tr{{ \text{tr}   }}
\def\Nul{{ \text{Nul}   }}
\def\Col{{ \text{Col}   }}

\def\be{{ \mathbf{e}  }}
\def\bx{{ \mathbf{x}  }}
\def\by{{ \mathbf{y}  }}

\newtheorem{theorem}{Theorem}
\newtheorem{lemma}{Lemma}
\newtheorem{remark}{Remark}
\newtheorem{corollary}{Corollary}

\begin{document}
\title{Secure MIMO Communications under Quantized Channel Feedback in the presence of Jamming }
\author{Theodoros Tsiligkaridis, \textit{Member, IEEE}}

\maketitle

\begin{abstract} 
We consider the problem of secure communications in a MIMO setting in the presence of an adversarial jammer equipped with $n_j$ transmit antennas and an eavesdropper equipped with $n_e$ receive antennas. A multiantenna transmitter, equipped with $n_t$ antennas, desires to secretly communicate a message to a multiantenna receiver equipped with $n_r$ antennas. We propose a transmission method based on artificial noise and linear precoding and a two-stage receiver method employing beamforming. Under this strategy, we first characterize the achievable secrecy rates of communication and prove that the achievable secure degrees-of-freedom (SDoF) is given by $d_s=n_r-n_j$ in the perfect channel state information (CSI) case. Second, we consider quantized CSI feedback using Grassmannian quantization of a function of the direct channel matrix and derive sufficient conditions for the quantization bit rate scaling as a function of transmit power for maintaining the achievable SDoF $d_s$ with perfect CSI and for having asymptotically zero secrecy rate loss due to quantization. Numerical simulations are also provided to support the theory.
\end{abstract}


\blfootnote{
Copyright (c) 2014 IEEE. Personal use of this material is permitted. However, permission to use this material for any other purposes must be obtained from the IEEE by sending a request to pubs-permissions@ieee.org.

The views expressed are those of the author and do not reflect the official policy or position of the Department of Defense or the U.S. Government.

Approved for Public Release, Distribution Unlimited.

T. Tsiligkaridis is with MIT Lincoln Laboratory, Lexington, MA 02421 USA (email: ttsili@ll.mit.edu).
}

\section{Introduction} \label{sec:intro}
Secrecy in the physical layer is concerned with maximizing the information rate of a transmitter-receiver pair such that reliable communication is possible, while keeping the information as private as possible if eavesdroppers listen. The seminal work of Wyner on the wiretap channel \cite{Wyner:1975} has shown that it is possible to reliably communicate at a strictly positive rate while an eavesdropper listening to the transmitted signal through its own channel cannot decode the message. These results were generalized to Gaussian channels by Leung-Yan-Cheong and Hellman in \cite{Cheong:1978} and to arbitrary broadcast channels by Csiszar and Korner in \cite{Csiszar:1978}. Following these important early works, various extensions to different system settings and assumptions have been made. Particularly, there has been considerable interest in studying physical layer secrecy for multiple-input multiple-output (MIMO) Gaussian channels, as the use of multiple antennas can increase secrecy capacity \cite{Khisti:2010:1, Khisti:2010:2}. In many works, the assumption of the transmitter knowing its channel to the eavesdropper is often impractical. As a result, transmission strategies exploiting multiple antennas and injecting controlled artificial noise into the transmitted signal have been proposed in \cite{Goel:2008} in order to enhance secrecy.

Quantized feedback schemes have been proposed and studied in the literature for single user and multiple user downlink communication systems in \cite{Jindal:2006, Yoo:2007}. Motivated by the quantization approach of Rezaee \& Guillaud \cite{Rezaee:2012, Rezaee:2013}, which considered interference alignment for the MIMO interference channel with quantized channel feedback, we study the value of quantized feedback for secrecy communications in the presense of a hostile jammer. A key motivator for our work is the paper by Krishnamachari et al. \cite{Krishnamachari:Aug:2013}, where a Grassmannian feedback scheme was studied in the context of interference alignment for MIMO interference channel. It was shown that if the feedback bit rate increases fast enough as a function of signal-to-noise ratio (SNR), the quantized channel estimates can be used at the transmitters to achieve the full multiplexing gain. However, communication under secrecy was not considered in these works on quantized CSI feedback. Optimal power allocation algorithms and achievable secrecy rates have been recently studied in the context of secrecy communications with artificial noise in \cite{Zhou:2010, Zhang:2013}, but no degree-of-freedom (DoF) analysis is performed and perfect CSI is assumed. To the best of our knowledge, the value of quantized channel feedback on the secure DoF gain has not been studied in the literature in the context of communicating under secrecy and in the presence of a hostile jammer.

The transmission model that we consider consists of transmitting a linear combination of artificial noise and desired signal to a receiver. The purpose of the artificial noise is to confuse the eavesdropper and the desired signal is the signal to be decoded by the intended receiver. The artificial noise aspect has been studied in \cite{Goel:2008, Lin:2011, Mukherjee:2011, Yang:2013}. In \cite{Goel:2008}, the approach of transmitting artificial noise in the nullspace of the direct channel matrix was proposed assuming the number of transmit antennas, $n_t$, are more than the number of receive antennas, $n_r$. Thus, the designed artificial noise has no impact on the received signal at the intended receiver, but causes degradation at the eavesdropper thus enhancing secrecy. Robust beamforming for secure MIMO communications was studied in \cite{Mukherjee:2011} under inaccurate CSI using a second-order perturbation analysis. The effect of delayed perfect CSIT on the SDoF gain was studied in \cite{Yang:2013} in the context of the two-user broadcast MIMO channel. In \cite{Lin:2011}, the optimal power allocation among artificial noise and desired signal is derived such that the ergodic secrecy rate is maximized for a fixed number of feedback bits and transmit power. In addition, a scaling law between feedback bits and power is derived to guarantee a constant secrecy rate loss compared to the perfect CSI case. Our work differs from these works since we derive explicit conditions for the quantization bit scaling that guarantee the optimal SDoF scaling, under similar assumptions on channel values as made in \cite{Krishnamachari:Aug:2013}, and in addition we consider a jammer interfering with the received signal and show that the secrecy rate loss due to quantization is asymptotically negligible as SNR grows. Furthermore, our work differs from \cite{Lin:2011} because we consider quantization of a function of the direct channel matrix using (deterministic) quantization theory on the Grassmann manifold, adopt different channel conditions as adopted in \cite{Rezaee:2012, Rezaee:2013, Krishnamachari:Aug:2013} and thus, develop different proof techniques. In addition, the SDoF performance of artificial noise transmission schemes has not been studied in \cite{Goel:2008}, and we also consider the lack of instantaneous perfect CSI by assuming the availability of quantized CSI. Our work differs from \cite{Mukherjee:2011, Yang:2013} in that we consider a jammer and the effect of Grassmannian-based quantized instantaneous CSI on SDoF gain in a point-to-point MIMO channel.

In our problem formulation, the receiver's strategy is to null out the jammer interference and use its remaining resources to recover the information-bearing signal from the transmitter through beamforming. Under perfect CSI conditions, assuming $n_t>n_r$, the transmitter designs the artificial noise signal to lie in the nullspace of the channel matrix $\bH_d$ and the information-bearing signal to lie in the orthogonal complement of the null space of $\bH_d$. Under this linear precoding strategy, the receiver sees no leakage due to artificial noise in the received signal. However, if the transmitter has imperfect CSI, then the received signal will contain a non-negligible amount of artificial noise. In the high SNR regime, this leakage will deteriorate the secrecy rate performance and drive the SDoF to zero. In order to maintain the full SDoF gain of the system, the rate of quantized feedback needs to increase appropriately as a function of SNR. In this paper, we characterize this rate and identify the key parameters associated with it. We also prove that, as the transmit power grows asymptotically, there is no loss in secrecy rate performance due to quantization.

\subsection{Outline}
The outline of this paper is as follows. Section \ref{sec:problem_formulation} formulates the basic problem. Section \ref{sec:rates_perfect_CSI} studies the achievable secrecy communication rates in the case of perfect CSI. Section \ref{sec:rates_quantized_CSI} studies the case of quantized CSI and derives performance bounds on the achievable secure degrees-of-freedom. The theory is illustrated by simulation in Section \ref{sec:simulations} and is followed by our conclusions in Section \ref{sec:conclusions}.

\subsection{Notation}
We use $\RR$ and $\CC$ to denote the real and complex fields. We use boldface lowercase letters $\bx$ to denote vectors and bold uppercase letters $\bA$ for matrices. Given a matrix $\bA\in \CC^{m\times n}$, we let $\bA^*$ denote its Hermitian conjugate, and $\bA^T$ denote its transpose. The trace operator $\tr(\cdot)$ on a square matrix is simply the sum of its diagonal entries. Let $\Nul(\bA)$ and $\Col(\bA)$ denote the nullspace and column spaces (i.e., range) of the matrix $\bA$.

We let $\mathcal{N}_c(\mathbf{\mu},\mathbf{\Sigma})$ denote the complex multivariate normal distribution with mean $\mathbf{\mu}$ and covariance $\mathbf{\Sigma}$. Consider two sequence of real numbers $\{a_P\}$ and $\{b_P\}$ indexed by $P$. Consider two sequences $\{a_P\}$ and $\{b_P\}$. The asymptotic notation $a_P=O(b_P)$ as $P\to\infty$ implies that there exists $K>0,P_0$ such that for all $P\geq P_0$, we have $|a_P|\leq K |b_P|$. The asymptotic notation $a_P=o(b_P)$ means that for all $\epsilon>0$, there exists $P_0(\epsilon)=P_0$ such that for all $P\geq P_0$, we have $|a_P|\leq \epsilon |b_P|$. We use $\log(\cdot)$ to denote the logarithm with base $2$ and $\log_e(\cdot)$ to denote the natural logarithm. Define the thresholding operator $(\cdot)_+=\max(\cdot,0)$.

\section{Problem Formulation} \label{sec:problem_formulation}
We consider the point-to-point MIMO channel with a transmitter (Tx) and a legitimate receiver (Rx). There is a jammer (J) degrading the received signal at Rx and an eavesdropper (Eve) observing a noisy version of the transmitted signal. Define the channel matrices $\bH_d\in \CC^{n_r\times n_t}$ as the channel between Tx and Rx, $\bH_e \in \CC^{n_e\times n_t}$ as the channel between the Tx and Eve and $\bH_j\in \CC^{n_r \times n_j}$ as the channel between J and the Rx. The received signals at the legitimate receiver and Eve are respectively given by:
\begin{align}
	\by       &= \bH_d \bx + \bH_j \bx_j + \bn & \text{(Rx)} \label{eq:Rx_y} \\
	\bar{\by} &= \bH_e \bx + \bar{\bn} & \text{(Eve)}          \label{eq:E_y}
\end{align}
where $\bx$ is the transmitted signal and $\bx_j$ is the jammer signal. The additive receiver noises $\bn$ and $\bar{\bn}$ are assumed to be white and Gaussian distributed, i.e., $\bn \sim \mathcal{N}_c(\mathbf{0},\sigma^2 \bI_{n_r}), \bar{\bn} \sim \mathcal{N}_c(\mathbf{0},\bar{\sigma}^2 \bI_{n_e})$. 

\begin{figure}[t]
	\centering
		\includegraphics[width=0.50\textwidth]{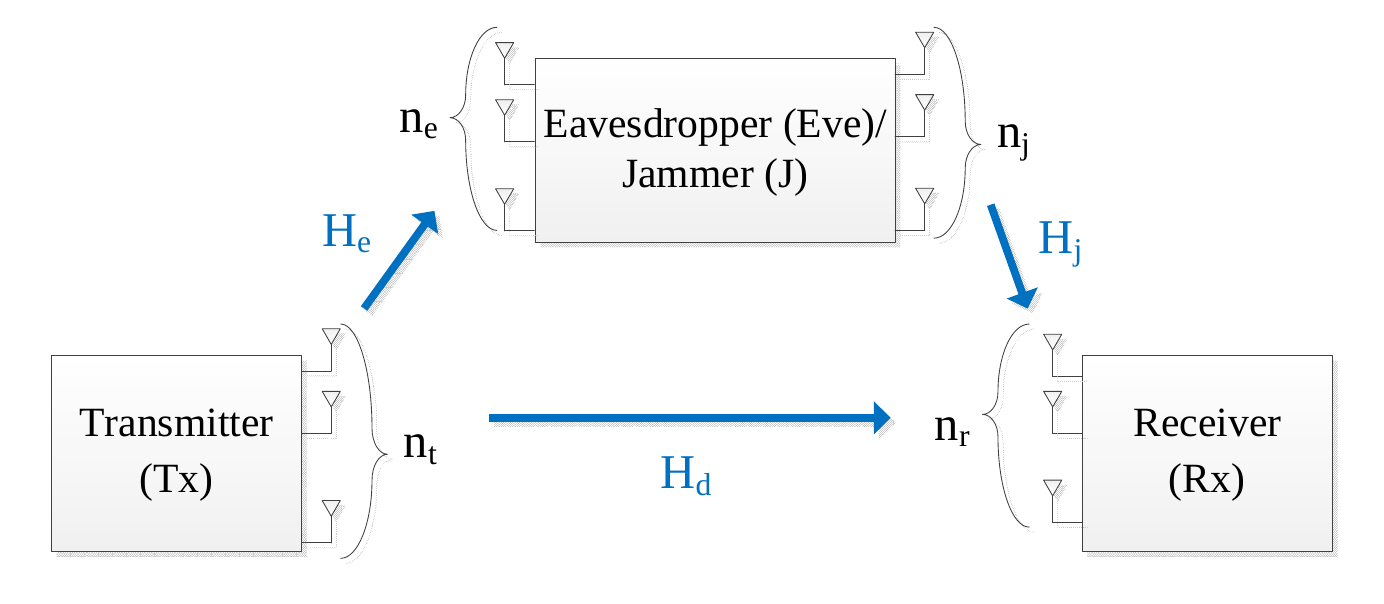}
	\caption{ Block diagram of secrecy communication problem in the presence of a jammer. }
	\label{fig:fig1}
\end{figure}

In this paper, we assume that Eve and J and cooperative in the sense that the jammer does not interfere with the eavesdropper signal. This can be realized in one of two ways. One way that this can be realized is if Eve and J are one simultaneous-transmit-and-receive unit, comprised of $n_e$ receive antennas and $n_j$ transmit antennas as depicted in Figure \ref{fig:fig1}. Another way that this can be realized is to have Eve and J be separate nodes with Eve having $N_e=n_e+n_j$ receive antennas and J having $n_j$ transmit antennas. Then, Eve can null out the undesired jammer interference by projecting its received signal in a subspace of dimension $n_e$, leading to the linear observation model (\ref{eq:E_y}). To illustrate this, say that Eve has $N_e$ receive antennas and observes:
\begin{equation*}
	\by_e = \bG_e \bx + \bG_j \bx_j + \bn_e
\end{equation*}
where $\bG_e$ is the channel from Tx to Eve, $\bG_j$ is the channel from J to Rx and $\bn_e$ is the receiver thermal noise at Eve. Considering the SVD of $\bG_j=\bU_j \bSigma_j \bV_j^*$, we can choose $\bU_{0,j} \in \CC^{N_e \times n_e}$ to consist of the columns of $\bU_j$ corresponding to zero singular values. Then, defining the projection $\bar{\by}=\bU_{0,j}^*\by_e$ \footnote{Here, $\Nul(\bU_{0,j}^*)=\Col(\bG_j)$ is the nullspace condition that is used in the nulling operation.}, we obtain the model (\ref{eq:E_y}) with $\bH_e=\bU_{0,j}^*\bG_e$ and $\bar{\bn}=\bU_{0,j}^*\bn_e$. Under the assumption that channel matrix elements are drawn i.i.d. from a continuous distribution, it follows that with probability 1, $\bH_e$ is full-rank and the equivalent model (\ref{eq:E_y}) can be used without loss of generality.

As in \cite{Krishnamachari:Aug:2013, Rezaee:2012, Rezaee:2013}, we assume all elements of the channel matrices are drawn i.i.d. from a continuous distribution, and the channel values do not change during the signal transmission. The receiver is assumed to have perfect knowledge of $\bH_d$ and $\bH_j$. As assumed in \cite{Krishnamachari:Aug:2013, Rezaee:2012, Rezaee:2013} we assume there is an error-free feedback link from the receiver to the transmitter. During the initial channel feedback phase, the receiver transmits its CSI using $N_f$ bits due to limited bandwidth. Data transmission then follows where the transmitter designs its signal using quantized feedback. 

\subsection{Transmitter Strategy}
Assuming $n_t>n_r$, the channel matrix $\bH_d$ has rank equal to $n_r$, which implies that there are $n_t-n_r$ dimensions available for artificial noise and $n_r$ dimensions for information transmission.

We assume the transmitter splits its available power $P$ into two components, the artificial noise and the information-bearing signal, using a linear combination:
\begin{equation} \label{eq:Tx}
	\bx = \sqrt{\rho} \bW_1 \bx_s + \sqrt{1-\rho} \bW_2 \bx_{an}
\end{equation}
where $\bx_s \in \CC^{n_r}$ is the information bearing signal, $\bx_{an} \in \CC^{n_t-n_r}$ is artificial noise and $\rho$ is the percentage of power allocated to $\bx_s$. The linear precoding matrices $\bW_1\in \CC^{n_t \times n_r}$ and $\bW_2\in \CC^{n_t\times n_t-n_r}$ are assumed to be truncated unitary matrices. We also assume that $\bx_s$ is independent of $\bx_{an}$. We let the covariance matrices of $\bx_s$ and $\bx_{an}$ be given by $\bK_{x_s} \in \CC^{n_r\times n_r}$ and $\bK_{x_{an}}\in \CC^{(n_t-n_r) \times (n_t-n_r)}$, respectively.

In order to design $(\bW_1,\bW_2)$ such that the artificial noise component cancels out at the receiver, the columns of $\bW_2$ must lie in the null space of $\bH_d \in \CC^{n_r\times n_t}$. This can be accomplished through the singular value decomposition (SVD) of $\bH_d$. Then, $\bH_d \bW_2=\mathbf{0}$.

\subsection{Receiver Strategy}
The receiver has instantaneous knowledge of $\bH_d$ and $\bH_j$. To null out the jammer's signal, the post-processing truncated unitary matrix $\bV \in \CC^{n_r \times n_r-n_j}$ at the receiver is applied:
\begin{equation} \label{eq:Rx_V}
	\tilde{\by} = \bV^*\mathbf{y} = \bV^*\bH_d\bx + \tilde{\bn} 
\end{equation}
where $\tilde{\bn}=\bV^*\bn \sim \mathcal{N}_c(\mathbf{0},\sigma^2 \bI)$ since $\bV^*\bV=\bI_{n_r-n_j}$. Consider the SVD of $\bH_j$:
\begin{align*}
	\bH_j &= \bU(\bH_j) \bSigma(\bH_j) \bV(\bH_j)^* \\
				&= \left[ \bU_{1}(\bH_j) | \bU_{0}(\bH_j) \right] \begin{bmatrix} \bSigma_{1}(\bH_j) \\ \mathbf{0}_{n_r-n_j\times n_j} \end{bmatrix} \bV(\bH_j)^*
\end{align*}
Choosing the columns of $\bV$ span the left nullspace of $\bH_j$, i.e.,
\begin{equation} \label{eq:V_nulling}
	\bV = \bU_0(\bH_j)
\end{equation}
yields the desired nulling condition $\bV^*\bH_j=\mathbf{0}$.

\subsection{Secrecy Degrees-of-Freedom (SDoF)}

The secrecy capacity is known as a quantity measuring the maximal rate of reliable communication while maintaining secrecy to Eve \cite{Csiszar:1978}. The secrecy capacity is generally known to be \cite{Csiszar:1978, Khisti:2010:2, Oggier:2011}:
\begin{equation*}
	C_s = \max_{\bK_{x_s}\succeq \mathbf{0},\tr(\bK_{x_s})\leq P} \left(I(\bx_s;\tilde{\by}) - I(\bx_s;\bar{\by})\right)_+
\end{equation*}
where $I(X;Y)$ denotes the mutual information between $X$ and $Y$ \cite{CoverThomas:2006}.

For a given input covariance matrix $\bK_{x_s}$ satisfying the power constraint $\tr(\bK_{x_s})\leq P$, the achievable secrecy rate under the models (\ref{eq:Tx}) and (\ref{eq:Rx_V}) is given by:
\begin{equation} \label{eq:Rs_def}
	R_s(\bK_{x_s}) = \left( I(\bx_s;\tilde{\by}) - I(\bx_s;\bar{\by}) \right)_+
\end{equation}
As Theorem 1 shows (see Appendix A), $R_s(\bK_{x_s})$ will be positive for $P$ large enough under certain assumptions, and thus the thresholding operator $(\cdot)_+$ can be omitted for the high SNR analysis to be presented in this paper. The (achievable) secure degrees-of-freedom (SDoF) are defined as:
\begin{equation*}
	d_s = \lim_{P\to\infty} \frac{R_s(\bK_{x_s})}{\log P}
\end{equation*}
The intuitive meaning of this metric is that there are $d_s$ data streams that can be reliably communicated to the receiver without having the eavesdropper being able to decode the transmitted information.

\section{Perfect CSI: Achievable Secrecy Rate \& SDoF} \label{sec:rates_perfect_CSI}

For the case of perfect CSI at the transmitter, the precoding matrix $\bW_2$ can be designed as $\bW_2=\bV_0(\bH_d) \in \CC^{n_t \times (n_t-n_r)}$, where the SVD of $\bH_d \in \CC^{n_r\times n_t}$ is given by
\begin{align*}
	\bH_d &= \bU(\bH_d) \bSigma(\bH_d) \bV(\bH_d)^* \\
				&= \bU(\bH_d) \left[ \bSigma_1(\bH_d) | \mathbf{0}_{n_r\times n_t-n_r} \right] \left[\bV_1(\bH_d) | \bV_0(\bH_d) \right]^*
\end{align*}
In that case, we can also choose $\bW_1=\bV_1(\bH_d) \in \CC^{n_t\times n_r}$ in order to guarantee orthogonality between the information bearing signal $\bW_1\bx_s$ and the artificial noise signal $\bW_2 \bx_{an}$. Then, the post-processed received signal at Rx and the received signal at Eve become:
\begin{align*}
	\tilde{\by} &= \sqrt{\rho} \bV^* \bH_d \bW_1 \bx_s + \tilde{\bn} \\
		  &= \sqrt{\rho} \bV^* \bU(\bH_d) \bSigma_1(\bH_d) \bV_1(\bH_d)^* \bV_1(\bH_d) \bx_s + \tilde{\bn} \\
			&= \sqrt{\rho} \bV^* \bU(\bH_d) \bSigma_1(\bH_d) \bx_s + \tilde{\bn} \\
			&= \sqrt{\rho} \bH \bx_s + \tilde{\bn} \\
	\bar{\by} &= \sqrt{\rho} \bH_e \bV_1(\bH_d) \bx_s + \sqrt{1-\rho} \bH_e \bV_0(\bH_d) \bx_{an} + \bar{\bn} \\
						&= \sqrt{\rho} \bH_{e,s} \bx_s + \sqrt{1-\rho} \bH_{e,an} \bx_{an} + \bar{\bn}
\end{align*}
where we defined $\bH = \bV^* \bU(\bH_d) \bSigma_1(\bH_d)$ as the transformed $(n_r-n_j) \times n_r$ channel matrix after post-processing. We also defined $\bH_{e,s}=\bH_e \bV_1(\bH_d)$ and $\bH_{e,an} = \bH_e \bV_0(\bH_d)$.

Using these expression for the received signals and the Gaussian noise assumptions, along with (\ref{eq:Rs_def}), we obtain an expression for the achievable secrecy rate under perfect CSI:
\begin{align*}
	&R_s(\bK_{x_s})= \log \det \left( \bI + \frac{\rho}{\sigma^2} \bH \bK_{x_s} \bH^* \right) \\
		  &- \log \left( \frac{\det(\rho \bH_{e,s} \bK_{x_s} \bH_{e,s}^* + \frac{(1-\rho)P}{n_t-n_r} \bH_{e,an} \bH_{e,an}^* + \bar{\sigma}^2 \bI)}{\det(\frac{(1-\rho)P}{n_t-n_r} \bH_{e,an} \bH_{e,an}^* + \bar{\sigma}^2 \bI)} \right)
\end{align*}
where we used $\bK_{x_s}=\Cov(\bx_s)$ and $\bK_{x_{an}}=\Cov(\bx_{an})=\frac{P}{n_t-n_r}\bI_{n_t-n_r}$.

The next theorem shows that $n_r-n_j$ SDoF are achievable under several mild assumptions.
\begin{theorem} \label{thm:perfect_CSI}
	Assume the following:
	\begin{itemize}
		\item there exists constants $c_0,c_1\in (0,1/n_r]$ such that
		\begin{equation} \label{eq:K_xs_bounds}
			c_0 P \bI \preceq \bK_{x_s} \preceq c_1 P \bI
		\end{equation}
		\item $n_t>n_r>n_j$
		\item $n_e \leq n_t-n_r$
	\end{itemize}
	In the case of perfect CSI, it is possible to achieve up to $d_s = n_r-n_j$ SDoF at most.
\end{theorem}
\begin{IEEEproof}
	See Appendix A.
\end{IEEEproof}
We remark that the SDoF $d_s$ is independent of $n_e$ because the eavesdropper's rate $I(\bx_s;\bar{\by})$ converges to a constant, dependent on $n_e$, and thus it has an asymptotically vanishing contribution to the SDoF, i.e., $\frac{I(\bx_s;\bar{\by})}{\log P}\to 0$ as $P\to\infty$.

The achievable $d_s=n_r-n_j$ secure DoF is optimal since the jammer signal lies in a $n_j$-dimensional space and to null it out without any knowledge of its intrinsic dimension, receive beamforming  leaves $n_r-n_j$ effective number of antennas at the receiver to use for successful decoding of the transmitted message.

The sufficient conditions of Thm. \ref{thm:perfect_CSI} are very mild, given our goal of achieving $n_r-n_j$ SDoF. If the condition $n_t>n_r$ is violated, the right nullspace of $\bH_d$ is empty, which implies that no precoding matrix $\bW_2$ exists such that $\bH_d \bW_2 = \mathbf{0}$. If $n_r>n_j$ is violated, then no post-processing matrix $\bV$ exists such that $\bV^*\bH_j=\mathbf{0}$, i.e., the receiver does not have enough antennas to cancel the jammer's signal. If $n_e> n_t-n_r$, then the eavesdropper has enough antennas to recover at least one data stream containing information because the artificial noise signal spans at most a $(n_t-n_r)$-dimensional space, leading to zero secrecy. The assumption $n_e +n_r \leq n_t$ was also made in Section VI in \cite{Lin:2011}. Thus, all assumptions are necessary to proceed.

\section{Quantized CSI: Achievable Secrecy Rate \& SDoF} \label{sec:rates_quantized_CSI}

When there is imperfect CSI knowledge at the transmitter, the condition $\bH_d \bW_2=\mathbf{0}$ will be violated and there will be some leakage of artificial noise at the receiver. As a consequence, the result of Theorem \ref{thm:perfect_CSI} no longer holds. In this section, we show that the same number of secure degrees-of-freedom can be achieved if the feedback rate scales fast enough as a function of transmit power.

\subsection{Quantization on the Grassmann manifold}
We assume that the receiver has perfect knowledge of the channel $\bH_d$. Thus, it can perform the QR decomposition of the conjugate transpose of the channel matrix, i.e., $\bH_d^*$:
\begin{equation} \label{eq:QR_decomposition}
	\bH_d^* = \bF\bC
\end{equation}
where $\bC \in \CC^{n_r\times n_r}$ is an invertible matrix and $\bF \in \CC^{n_t \times n_r}$ is a tall orthonormal matrix whose columns span the same column space of $\bH_d^*$. Thus, from the invertibility of $\bC$, the condition $\bH_d\bW_2=\mathbf{0}$ is equivalent to $\bF^* \bW_2 = \mathbf{0}$. A similar decomposition approach was considered in \cite{Rezaee:2013} for the MIMO interference channel. As in \cite{Rezaee:2013}, a quantizer at the receiver uses $N_f$ bits to describe the columns of $\bF$ and transmits the index of the quantized codeword back to the transmitter through a noiseless feedback link. The transmitter and receiver share a predefined codebook $\mathcal{S} = \{\bS_1,\dots,\bS_{2^{N_f}}\}$ consisting of truncated unitary matrices of size $n_t\times n_r$, which is designed using Grassmannian subspace packing. The quantization of the matrix $\bF$ on the Grassmann manifold $\mathcal{G}_{n_t,n_r}$ is mathematically described by the minimum distance problem:
\begin{equation} \label{eq:F_design}
	\hat{\bF} = \arg \min_{\bS\in \mathcal{S}} d_c(\bS,\bF)
\end{equation}
where $d_c(\bS,\bF)=\frac{1}{\sqrt{2}} \nn\bS\bS^* - \bF\bF^*\nn_F$ is the chordal distance between $\bS$ and $\bF$ in $\mathcal{G}_{n_t,n_r}$.

\subsection{Transmitter Strategy under Quantized CSI}
Given the quantized matrix $\hat{\bF} \in \CC^{n_t\times n_r}$, the transmitter designs the linear precoding matrices $\bW_{1,Q}$ and $\bW_{2,Q}$. Let the matrix $\bW_{2,Q}\in \CC^{n_t\times n_t-n_r}$ be chosen such that
\begin{equation} \label{eq:F_constraint}
	\hat{\bF}^* \bW_{2,Q} = \mathbf{0}_{n_r\times n_t-n_r}
\end{equation}
This can be accomplished if the columns of $\bW_{2,Q}$ are chosen to span the nullspace of $\hat{\bF}^*$. We thus let $\bW_{2,Q}$ be chosen such that its columns form a basis for $\Nul(\hat{\bF}^*)$.

In order to maximize the amount of information being sent over the noisy channel $\bH_d$, the precoded information signal $\bW_{1,Q}\bx_s$ must always be orthogonal to the precoded artificial noise signal $\bW_{2,Q}\bx_{an}$. In order for this to hold irrespective of the signals $\bx_s$ and $\bx_{an}$, $\bW_{1,Q}$ must be orthogonal to $\bW_{2,Q}$. This orthogonality implies that the matrix $\bW_{1,Q} \in \CC^{n_t\times n_r}$ must lie in the orthogonal complement of $\Nul(\hat{\bF}^*)$. In other words, we let the columns of $\bW_{1,Q}$ form a basis for $\Col(\hat{\bF})$, i.e, $\bW_{1,Q}=\hat{\bF}$.

\subsection{Receiver Strategy under Quantized CSI}
Given $\by$, the receiver uses a post-processing matrix to form the transformed vector:
\begin{equation} \label{eq:y_postprocessed}
	\check{\by} = \bG^*\bV^*\by
\end{equation}
where $\bV$ is the nulling matrix for the jammer defined in (\ref{eq:V_nulling}) and $\by$ is the received signal in (\ref{eq:Rx_y}). Given that $\bV$ is already chosen using (\ref{eq:V_nulling}), we want to design $\bG\in \CC^{d_s \times d_s}$ as a function of $\bV,\bF,\bC$ (all of which are available at the receiver). Let us choose $\bG$ as:
\begin{equation} \label{eq:G_def}
	\bG^* = \bB^* \bF \bC \bV (\bV^* \bC^* \bC \bV)^{-1}
\end{equation}
where $\bB\in \CC^{n_t\times d_s}$ is a tall truncated unitary matrix of full rank.

Let the leakage term be given by 
\begin{equation*}
	\be_L = \sqrt{1-\rho} \bG^* \bV^* \bH_d \bW_{2,Q} \bx_{an}
\end{equation*}
The post-processed received signal can be written as
\begin{align*}
	\check{\by} &= \bG^*(\bV^*\bH_d\bx + \tilde{\bn}) \\
		&= \sqrt{\rho} \bG^*\bV^* \bH_d \bW_{1,Q} \bx_s + \be_L + \bG^*\tilde{\bn}
\end{align*}

\subsection{Controlling the Leakage Power}
The next lemma bounds the power of the leakage term.
\begin{lemma} \label{lemma:leakage_power}
	Assuming a quantization codebook construction based on sphere-packing using $N_f$ bits, the leakage power is bounded as:
	\begin{equation} \label{eq:L_bnd}
		L(P) := \EE \nn \be_L \nn_2^2 \leq \frac{2(1-\rho) P}{n_t-n_r} \left(\frac{2}{(c 2^{N_f})^{1/N}}\right)^2 (1+o(2^{-N_f/N}))
	\end{equation}
	where $N=2n_r(n_t-n_r)$ and $c$ is a constant.
\end{lemma}
\begin{IEEEproof}
	See Appendix B.
\end{IEEEproof}

The corollary that follows provides a sufficient condition on the feedback rate to ensure that the leakage power stays bounded. This will be a key condition that will be used to prove the optimal SDoF scaling.
\begin{corollary} \label{cor:L_bounded}
	Assume that the quantization with $N_f$ feedback bits scales as:
	\begin{equation} \label{eq:Nf_scaling}
		N_f = \frac{N}{2} \log_2 P = n_r(n_t-n_r) \log_2 P
	\end{equation}
	Then, the leakage power $L(P)$ stays bounded as $P\to\infty$, i.e., $L(P)=O(1)$ as $P\to\infty$.
\end{corollary}
\begin{IEEEproof}
	See Appendix C.
\end{IEEEproof}
We note that boundedness of the leakage power does not necessarily imply the achievability of the optimal SDoF. However, it provides a grasp on the feedback bit rate scaling that can possibly guarantee such a claim. Next, we show that the scaling condition (\ref{eq:Nf_scaling}) is sufficient to guarantee the optimal SDoF gain.

\subsection{SDoF Analysis}
Here, we show that the secrecy rate of the quantized CSI scheme achieves the same SDoF as the corresponding scheme with perfect CSI as derived in Theorem \ref{thm:perfect_CSI}. In other words, there is no performance loss for large enough SNR.

Under the transmitter and receiver strategies proposed for quantized CSI, the achievable secrecy rate when CSI is perfect is given in (\ref{eq:RsP}) by:
\begin{equation*}
	R_{s,G}^P = R_{s,G}^P(\bK_{x_s}) = I(\bx_{s};\check{\by})-I(\bx_s;\bar{\by}),
\end{equation*}
where the subscript $s$ denotes secrecy and the subscript $G$ denotes post-processing with the matrix $\bG^*$ in addition to $\bV^*$ (recall (\ref{eq:y_postprocessed})). Similarly, when only quantized CSI is available, the achievable secrecy rate is given by $R_{s,G}^Q$ in (\ref{eq:RsQ}), where $(\bW_{1,Q},\bW_{2,Q})$ are the designed precoding matrices under imperfect CSI. We note that in general $\bW_1\neq \bW_{1,Q}$ and $\bW_2\neq \bW_{2,Q}$, although they have the same rank.

\begin{figure*}[!t]
\normalsize
\begin{equation} \label{eq:RsP}
	R_{s,G}^P = \underbrace{\log\left( \frac{\det(\rho \bG^*\bV^*\bH_d \bW_1 \bK_{x_s} \bW_1^* \bH_d^*\bV\bG + \sigma^2 \bG^*\bG)}{\det(\sigma^2 \bG^*\bG)} \right)}_{T_{+}^P} - \underbrace{\log\left( \frac{\det(\rho \bH_e \bW_1 \bK_{x_s} \bW_1^* \bH_e^* + \frac{(1-\rho)P}{n_t-n_r} \bH_e\bW_2\bW_2^*\bH_e^* + \bar{\sigma}^2 \bI_{n_e} )}{\det(\frac{(1-\rho)P}{n_t-n_r} \bH_e\bW_2\bW_2^*\bH_e^* + \bar{\sigma}^2 \bI_{n_e})} \right)}_{T_{-}^P}
\end{equation}
\begin{align}
	R_{s,G}^Q &= \underbrace{\log\left( \frac{\det(\rho\bG^*\bV^*\bH_d\bW_{1,Q} \bK_{x_s} \bW_{1,Q}^* \bH_d^* \bV \bG + \frac{(1-\rho)P}{n_t-n_r} \bG^* \bV^* \bH_d \bW_{2,Q} \bW_{2,Q}^* \bH_d^*\bV\bG + \sigma^2 \bG^*\bG )}{\det(\frac{(1-\rho)P}{n_t-n_r} \bG^* \bV^* \bH_d \bW_{2,Q} \bW_{2,Q}^* \bH_d^*\bV\bG + \sigma^2 \bG^*\bG )} \right)}_{T_{+}^Q} \nonumber \\
		&\qquad - \underbrace{\log\left( \frac{\det(\rho \bH_e \bW_{1,Q} \bK_{x_s} \bW_{1,Q}^* \bH_e^* + \frac{(1-\rho)P}{n_t-n_r} \bH_e\bW_{2,Q}\bW_{2,Q}^*\bH_e^* + \bar{\sigma}^2 \bI_{n_e} )}{\det(\frac{(1-\rho)P}{n_t-n_r} \bH_e\bW_{2,Q}\bW_{2,Q}^*\bH_e^* + \bar{\sigma}^2 \bI_{n_e})} \right)}_{T_{-}^Q} \label{eq:RsQ}
\end{align}
\hrulefill
\vspace*{4pt}
\end{figure*}

Before proving the main result, we will need a few technical lemmas. We first recall the variational representation of the log-determinant function.
\begin{lemma} \cite{Bertsekas:1999} \label{lemma:logdet}
	Let $\bE\in \CC^{n\times n}$ be a positive definite matrix. Then,
	\begin{equation*}
		\log_e \det(\bE^{-1}) = \max_{\bS \succeq \mathbf{0}} \left\{-\tr(\bS\bE) + \log_e \det(\bS) + n\right\}
	\end{equation*}
	and the optimal solution is $\bS^*=\bE^{-1}$.
\end{lemma}

Lemma \ref{lemma:logdet} implies the following perturbation bounds, which will be crucial for analyzing the secrecy rate performance.
\begin{lemma} \label{lemma:logdet_perturbation}
	Let $\bA$ and $\bA + \bDelta$ be positive definite matrices. Then, the following bounds hold:
	\begin{align*}
		\log_e \det(\bA+\bDelta) - \log_e \det(\bA) &\leq \tr(\bA^{-1} \bDelta) \\
		\log_e \det(\bA+\bDelta) - \log_e \det(\bA) &\geq \tr(\bDelta (\bA+\bDelta)^{-1})
	\end{align*}
\end{lemma}

We will need a technical lemma that yields an asymptotic lower bound to a remainder term that will be crucial for proving the main result of the paper (Theorem \ref{thm:quantized_CSI}). 
\begin{lemma} \label{lemma:beta_P}
Consider the transmitter and receiver strategies described in Section \ref{sec:rates_quantized_CSI}. Assume the same conditions as in Theorem \ref{thm:perfect_CSI}. In addition, assume (\ref{eq:Nf_scaling}). Define the auxiliary variable $\beta(P)$ as:
\begin{align*}
	&\beta(P) = \log\det\Big( \rho P \bG^*\bV^*\bH_d \bW_{1,Q}\bW_{1,Q}^* \bH_d^* \bV\bG \\
	&\quad + (1-\rho)\frac{P}{n_t-n_r} \bG^*\bV^*\bH_d\bW_{2,Q}\bW_{2,Q}^*\bH_d\bV\bG + \sigma^2 \bG^*\bG \Big) \\
	&\quad - \log\det\left( \rho P \bG^*\bV^*\bH_d \bW_{1,Q}\bW_{1,Q}^* \bH_d^* \bV\bG + \sigma^2 \bG^*\bG  \right)
\end{align*}
Then, there exists a non-negative sequence $\{\epsilon(P)\}$ converging to zero such that for all $P$:
\begin{equation*}
	\beta(P) \geq -\epsilon(P)
\end{equation*}
\end{lemma}	
\begin{IEEEproof}
	See Appendix D.
\end{IEEEproof}

The main result of the paper is provided in Theorem \ref{thm:quantized_CSI}, where the optimal SDoF is shown to be obtained under the quantized feedback of CSI. In addition, it is shown that there is no secrecy rate loss due to quantization asymptotically as $P\to\infty$.
\begin{theorem} \label{thm:quantized_CSI}
	Consider the transmitter and receiver strategies described in Section \ref{sec:rates_quantized_CSI}. Assume the same conditions as in Theorem \ref{thm:perfect_CSI}. 
	\begin{enumerate}
	\item Assuming (\ref{eq:Nf_scaling}), i.e.,
	\begin{equation*}
		N_f = \frac{N}{2} \log_2 P
	\end{equation*}
	then, the full $d_s^Q=d_s=n_r-n_j$ SDoF are achievable.
	\item Assuming for some $\epsilon>0$,
	\begin{equation} \label{eq:epsilon_Nf}
		N_f = (1+\epsilon) \frac{N}{2} \log_2 P
	\end{equation}
	then, the full $d_s^Q=d_s=n_r-n_j$ SDoF are achievable and the asymptotic rate gap due to quantization $\delta_{\text{GAP}}:=\lim_{P\to\infty} \{R_{s,G}^P-R_{s,G}^Q\}$ is zero. 
	\end{enumerate}
\end{theorem}
\begin{IEEEproof}
	See Appendix E.
\end{IEEEproof}
\begin{remark}
	The scaling condition (\ref{eq:epsilon_Nf}) can be weakened to
	\begin{equation*}
		N_f = \frac{N}{2} \log_2(P m(P)) 
	\end{equation*}
	where $m(P)$ is any nonnegative function satisfying $m(P)\to\infty$ as $P\to\infty$.
\end{remark}

We finally remark that having $N_f$ be a monotonically increasing function in $P$ is not enough to guarantee $\delta_{\text{GAP}}=0$. This can be seen from the proof of Theorem \ref{thm:quantized_CSI}, where in order for $\delta_{\text{GAP}}$ to converge to zero as $P\to\infty$, we need the following condition to hold:
\begin{equation*}
	\bU(P) = P \cdot \bG^*\bV^*\bH_d \bW_{2,Q}\bW_{2,Q}^*\bH_d^*\bV\bG \to \mathbf{0}
\end{equation*}
which implies that $\bG^*\bV^*\bH_d \bW_{2,Q}\bW_{2,Q}^*\bH_d^*\bV\bG=o(P^{-1})$. The condition $N_f\to \infty$ alone only implies $\bG^*\bV^*\bH_d \bW_{2,Q}\bW_{2,Q}^*\bH_d^*\bV\bG=o(1)$ and does not guarantee $\bU(P)\to \mathbf{0}$. Necessary conditions that guarantee the results of Theorem \ref{thm:quantized_CSI} remain an open problem.

\section{Simulations} \label{sec:simulations}
This section contains a few illustrative simulations that validate the methodology presented throughout the paper. The elements of all channels were generated as i.i.d. random complex-normal $\mathcal{N}_c(0,1)$ random variables as in \cite{Rezaee:2012, Rezaee:2013}.

\begin{figure*}
    \centering
    \subfloat{\includegraphics[width=0.48\textwidth]{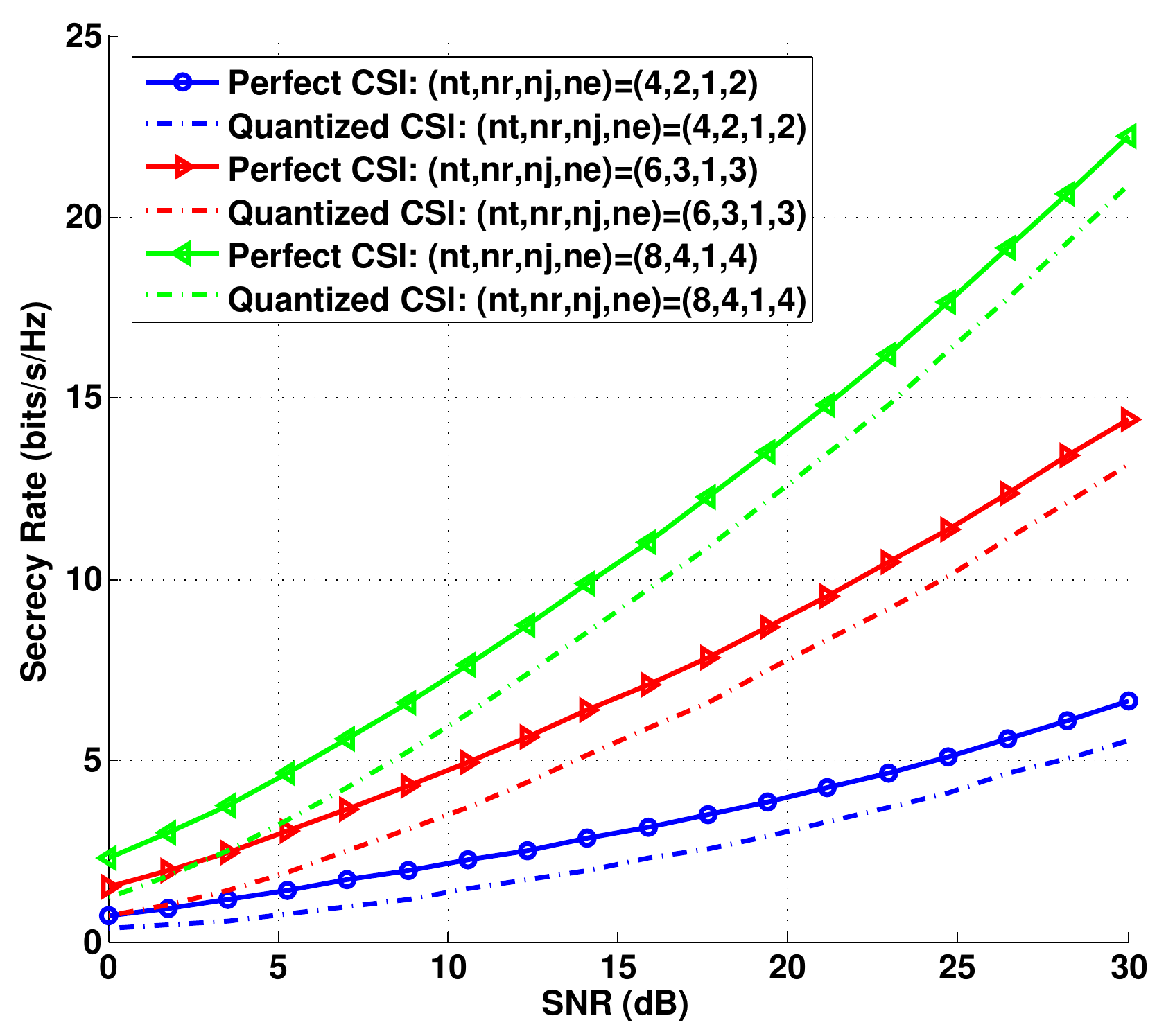}}
    \qquad
    \subfloat{\includegraphics[width=0.48\textwidth]{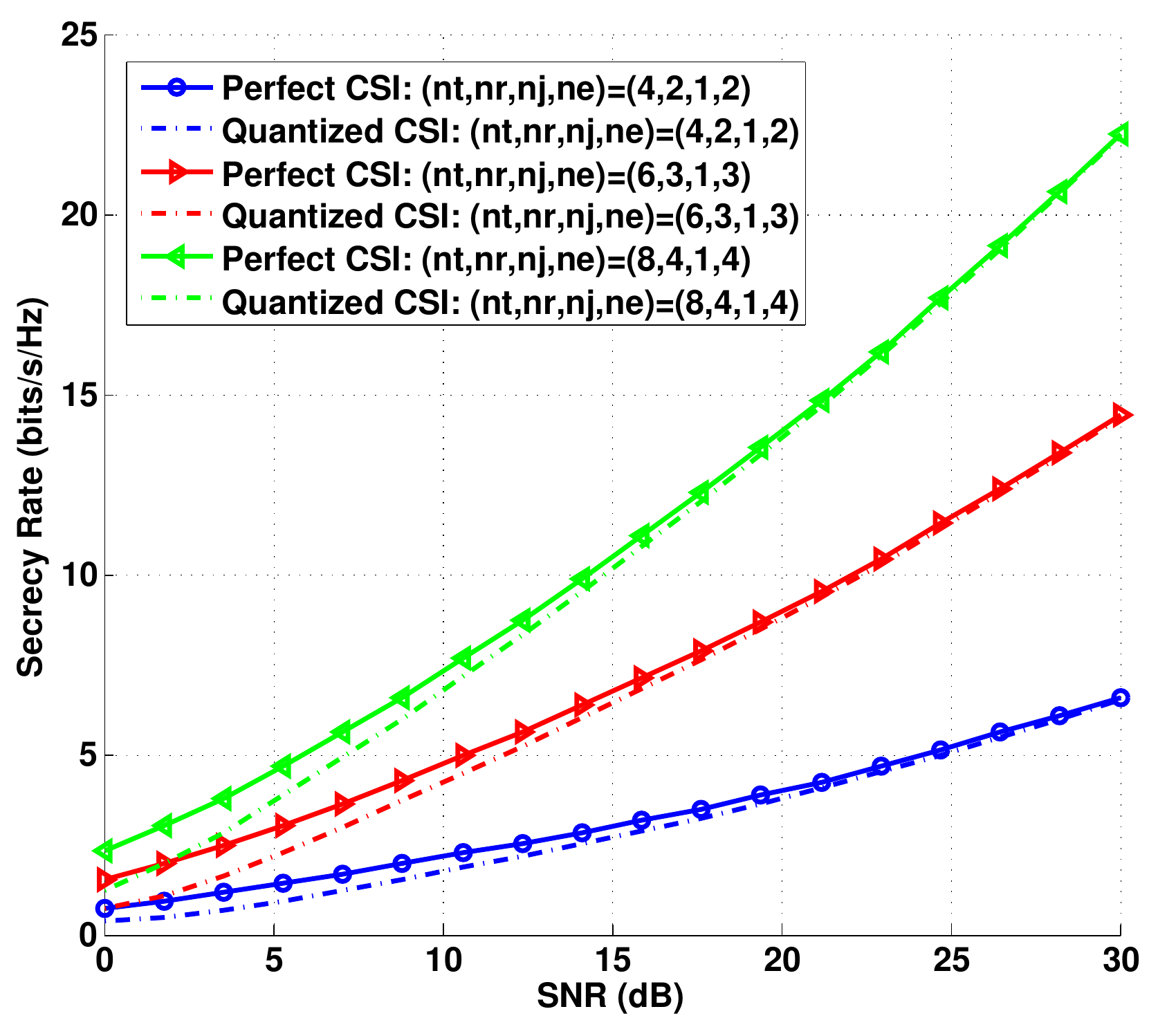}}
    \caption{\small{Monte Carlo simulation for secrecy rate performance as a function of $SNR$ for $N_f=\frac{N}{2}\log_2 P$ (left panel) and $N_f=\frac{(1+0.5)N}{2}\log_2 P$ (right panel). The transmitter has $n_t=2n_r$ antennas and the receiver has $n_r=2,3,4$ antennas. The jammer is equipped with one antenna, i.e., $n_j=1$, and the eavesdropper is equipped with $n_e=n_t-n_r=n_r$ antennas. Equal power allocation $\rho=1/2$ was used. For high SNR, the slope of the secrecy rate curves corresponding to quantized CSI become identical to the slope of the curves with perfect CSI, as predicted by Theorem \ref{thm:quantized_CSI}. The corresponding slopes are $d_s=1,2,3$. As predicted in Theorem \ref{thm:quantized_CSI}, the secrecy rate gap converges to zero as $P\to\infty$ since $N_f$ increases slightly faster than prescribed in (\ref{eq:Nf_scaling}) (see right panel). } }
    \label{fig:secrecy_rate1}
\end{figure*}

Figure \ref{fig:secrecy_rate1} shows the secrecy rate performance as a function of transmit SNR for $n_r=2,3,4$ where $n_j=1$ and $n_t=2n_r, n_e=n_t-n_r=n_r$. The feedback bit rate increases as a function of $P$ according to (\ref{eq:Nf_scaling}) for the left panel and according to (\ref{eq:epsilon_Nf}) for the right panel. The secrecy rate for the perfect CSI and quantized CSI schemes were calculated using the expressions in (\ref{eq:RsP}) and (\ref{eq:RsQ}) with the choices $\bK_{x_s}=\frac{P}{n_r}\bI_{n_r}$ and $\bK_{x_an}=\frac{P}{n_t-n_r}\bI_{n_t-n_r}$. Due to the high complexity associated with implementing (\ref{eq:F_design}) for large bit rates $N_f$, we adopt the random perturbation scheme of \cite{Rezaee:2013} to generate the quantized matrices $\hat{\bF}$\footnote{The reason why $n_t\geq 2n_r$ is chosen here has to do with the approximation of the quantization. While not pursued in this paper, the case $n_t<2n_r$ can be covered in a similar manner. For more details, see Section VI.B in \cite{Rezaee:2013}.}. It was shown numerically in \cite{Rezaee:2013} that this approximation is fairly accurate for a wide range of SNR. It is evident from Figure \ref{fig:secrecy_rate1} that the slopes of the secrecy rate curves become identical as SNR grows, implying that the SDoF become identical, as expected from Theorem \ref{thm:quantized_CSI}. In addition, for $N_f=\frac{(1+\epsilon)N}{2}\log_2 P$ with $\epsilon=0.5$, the right panel of Figure \ref{fig:secrecy_rate1} shows that the secrecy rate gap covnerges to zero as $P$ grows to infinity.

\begin{figure}[ht]
	\centering
		\includegraphics[width=0.48\textwidth]{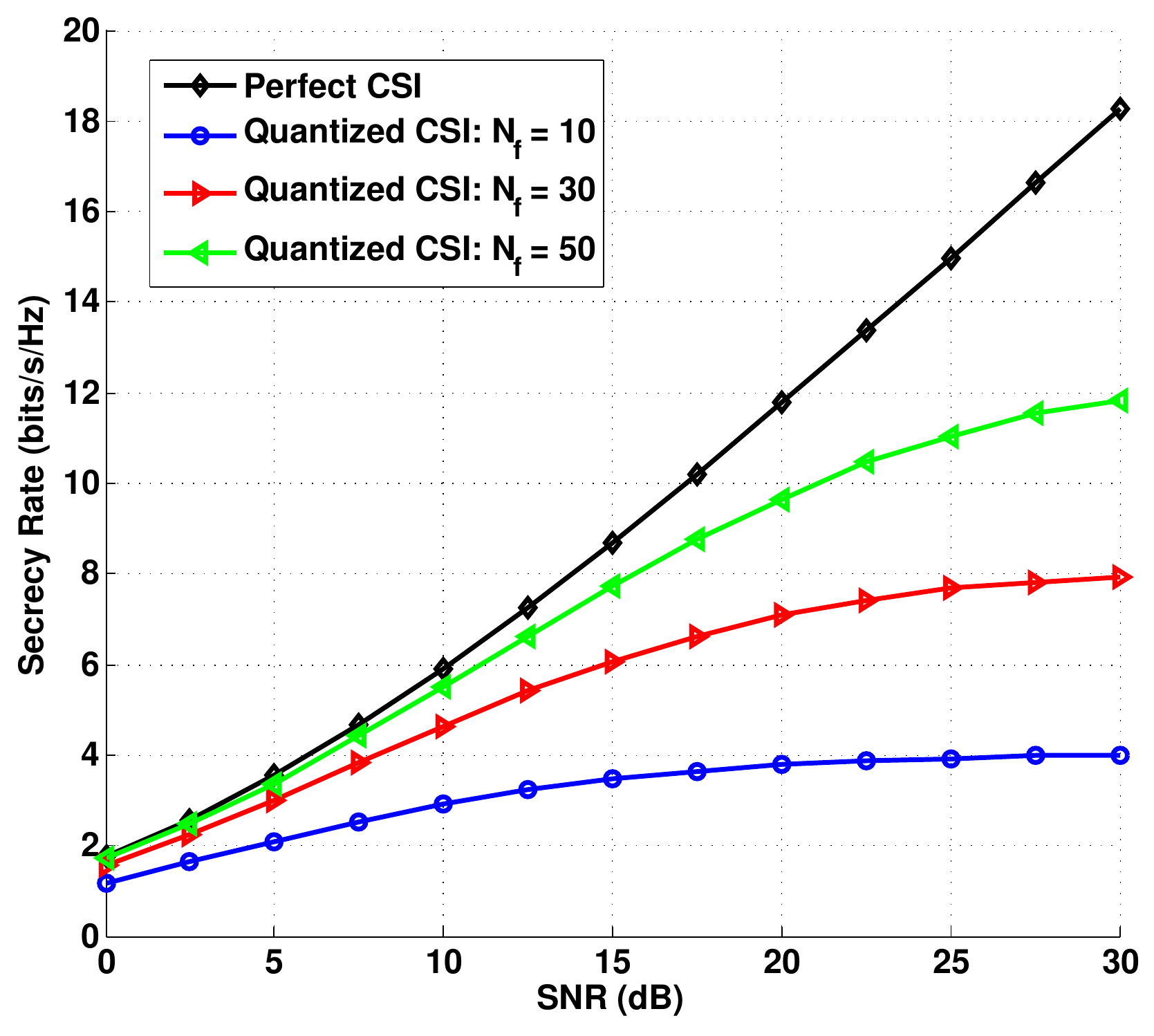}
	\caption{\small{Monte Carlo simulation for secrecy rate loss due to quantization as a function of $SNR$ for $N_f=30,60,90$. The transmitter has $n_t=2n_r$ antennas and the receiver has $n_r=3$ antennas. The eavesdropper has $n_e=n_t-n_r=3$ antennas and the eavesdropper has a single antenna. Equal power allocation $\rho=1/2$ was used. The secrecy rate saturates to a limiting value as SNR grows, implying zero SDoF gain.}  }
	\label{fig:secrecy_rate2}
\hfill
	\centering
		\includegraphics[width=0.48\textwidth]{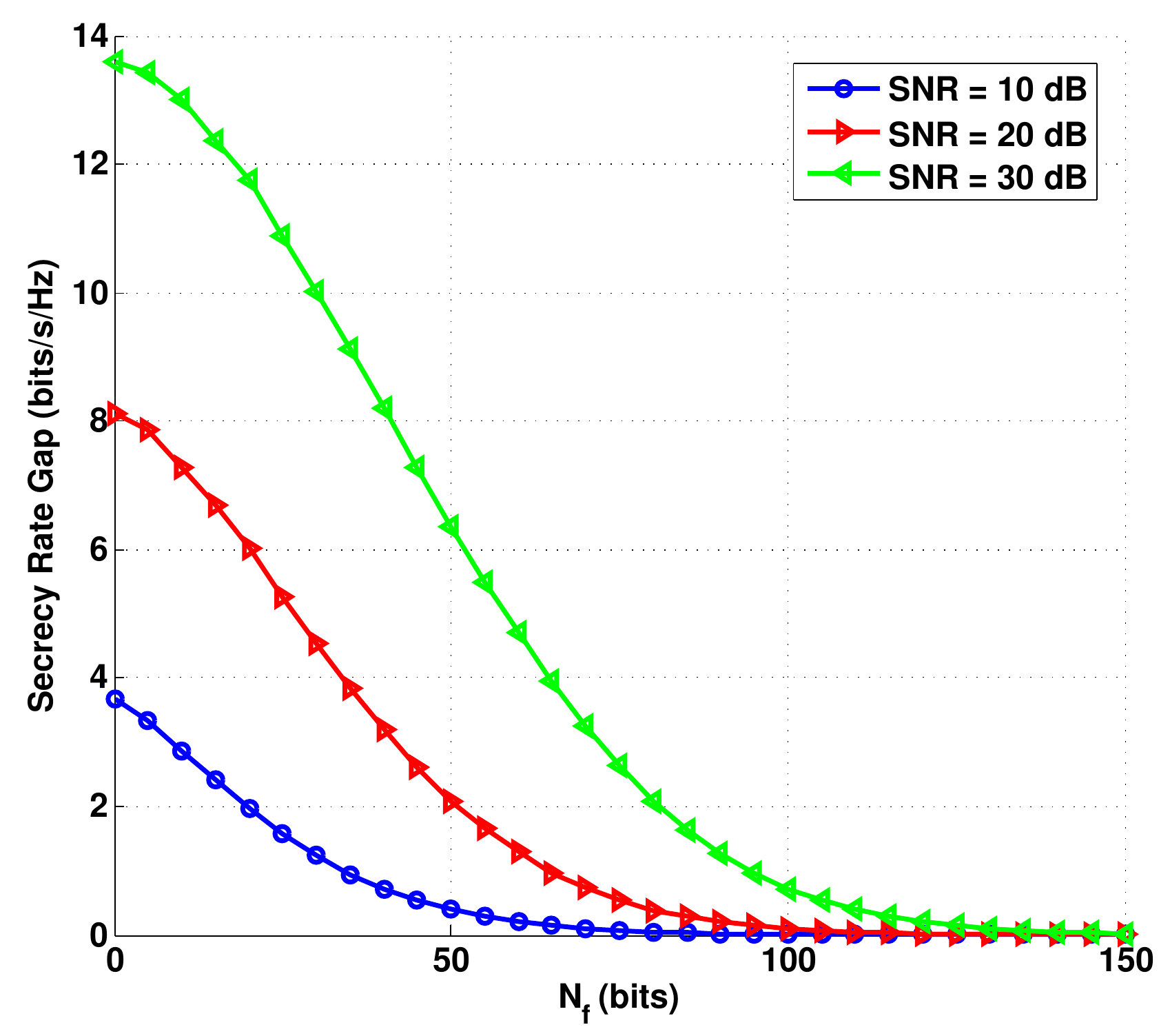}
	\caption{\small{Monte Carlo simulation for secrecy rate loss due to quantization as a function of $N_f$ for fixed $SNR$. The transmitter has $n_t=2n_r$ antennas and the receiver has $n_r=3$ antennas. The jammer has $n_j=1$ antenna and the eavesdropper has $n_e=n_t-n_r=3$ antennas. Equal power allocation $\rho=1/2$ was used. The secrecy rate gap converges to zero fast as the number of feedback bits increase.}  }
	\label{fig:secrecy_rate3}
\end{figure}

Figure \ref{fig:secrecy_rate2} shows the secrecy rate performance as a function of SNR for $n_r=3$ where $n_t=2n_r$ and $n_e=n_t-n_r$. The feedback bit rate is fixed to $N_f=30,60,90$ and it is observed that the secrecy rate converges to a limiting value as the transmit SNR grows to infinity. Similar behavior is observed for the communication rate performance with finite rate feedback without secrecy or jamming in \cite{Rezaee:2012, Rezaee:2013, Jindal:2006, Yoo:2007} and without jamming in \cite{Lin:2011}. This phenomenon is due to the fact that finite quantization bit rate allows the artificial noise to dominate at the receiver and as a result, the secure multiplexing gain becomes zero.

Figure \ref{fig:secrecy_rate3} shows the loss in secrecy rate due to quantization, for fixed SNR and number of antennas, as a function of number of feedback bits $N_f$. The SNR is fixed to $SNR=10,20,30$ dB and we observe that the secrecy rate loss converges to zero fast as $N_f$ grows. We conjecture that this rate gap converges to zero exponentially fast as a function of $N_f$.

\section{Conclusion} \label{sec:conclusions}
We studied the value of quantized feedback for MIMO secrecy communications in the presence of a jammer. We proposed transmitter and receiver strategies based on linear precoding and receive beamforming to simultaneously combat jammer interference, imperfect CSI at the transmitter and eavesdropping. Under this MIMO communication model, we characterized the achievable secrecy rate performance of the system. We derived sufficient conditions on the feedback bit rate scaling as a function of transmit power that guarantees the same secure degrees-of-freedom as the corresponding scheme with perfect CSI. We also showed that there is no secrecy rate loss due to quantization asymptotically as $P\to\infty$ under the same conditions. Simulations were shown to validate the theoretical analysis.

Future work may include deriving necessary conditions on the feedback bit rate to maintain the optimal secure degrees-of-freedom and developing efficient power allocations algorithms to improve performance in the low SNR regime. Another worthwhile open problem is to bound the secrecy rate gap as a function of finite power $P$ and number of feedback bits $N_f$. This type of analysis would provide insight into how many feedback bits are needed to achieve an arbitrarily small secrecy rate gap.

\section*{Acknowledgment}
The author gratefully acknowledges discussions with David Browne and Keith W. Forsythe. 

\appendices

\section{Proof of Theorem \ref{thm:perfect_CSI}}

\begin{IEEEproof}
  Decompose $R_s$ into two components $R_s = R_s^d - R_s^E$	where
	\begin{align*}
		R_s^d &= \log \det \left( \bI + \frac{\rho}{\sigma^2} \bH \bK_{x_s} \bH^* \right) \\
		R_s^E &= \log \left( \frac{\det(\rho \bH_{e,s} \bK_{x_s} \bH_{e,s}^* + \frac{(1-\rho)P}{n_t-n_r} \bH_{e,an}\bH_{e,an}^* + \bar{\sigma}^2 \bI)}{\det(\frac{(1-\rho)P}{n_t-n_r} \bH_{e,an}\bH_{e,an}^* + \bar{\sigma}^2 \bI)} \right)
	\end{align*}
	Expanding $R_s^d$, we obtain:
	\begin{align}
		R_s^d &= \sum_{i=1}^{n_r-n_j} \log\left(1 + \frac{\rho}{\sigma^2} \lambda_i(\bH\bK_{x_s}\bH^*) \right) \nonumber \\
				  &= d_s \log(P) + \sum_{i=1}^{d_s}	\log\left(P^{-1} + \frac{\rho}{\sigma^2} \frac{\lambda_i(\bH\bK_{x_s}\bH^*)}{P} \right) \label{eq:e1}
	\end{align}
	Using (\ref{eq:K_xs_bounds}), it is guaranteed that all eigenvalues $\lambda_i(\bH\bK_{x_s}\bH^*)$ satisfy the bounds:
	\begin{equation} \label{eq:lambda_bounds}
		c_0 P \lambda_i(\bH\bH^*) \leq \lambda_i(\bH\bK_{x_s}\bH^*) \leq c_1 P \lambda_i(\bH\bH^*)
	\end{equation}
	Since the matrix $\bH\bH^*$ is positive definite almost surely, using (\ref{eq:lambda_bounds}) into (\ref{eq:e1}) and taking the limit as $P\to\infty$, we obtain $\frac{R_s^d}{\log(P)} \stackrel{P\to\infty}{\longrightarrow} d_s$. To finish the proof, it remains to show $\frac{R_s^E}{\log(P)} \stackrel{P\to\infty}{\longrightarrow} 0$. To prove this, note that $R_s^E$ converges to a constant as $P\to\infty$:
	\begin{align*}
		&R_s^E \\
			&= \log\det\left(\bI + \rho \bH_{e,s} \bK_{x_s} \bH_{e,s}^* \left( \frac{(1-\rho)P\bH_{e,an}\bH_{e,an}^* }{n_t-n_r} + \bar{\sigma}^2 \bI \right)^{-1} \right) \\
			&\leq \log\det\left( \bI + c_1 \rho P \bH_{e,s} \bH_{e,s}^* \left( \frac{(1-\rho)P\bH_{e,an}\bH_{e,an}^*}{n_t-n_r}  + \bar{\sigma}^2 \bI \right)^{-1} \right) \\
			&= \log\det\left( \bI + c_1 \rho \bH_{e,s} \bH_{e,s}^* \left( \frac{(1-\rho)\bH_{e,an} \bH_{e,an}^*}{n_t-n_r}  + \bar{\sigma}^2 P^{-1} \bI \right)^{-1} \right) \\
			&\stackrel{P\to\infty}{\to} \log\det\left( \bI + c_1 \frac{\rho(n_t-n_r)}{1-\rho} \bH_{e,s} \bH_{e,s}^* \left( \bH_{e,an} \bH_{e,an}^* \right)^{-1} \right) \\
	\end{align*}
	Note that we used the fact that $n_e\leq n_t-n_r$, implying that $\bH_{e,an}\bH_{e,an}^*$ is full rank. Using a similar argument and (\ref{eq:K_xs_bounds}) again, it follows that $R_s^E$ is bounded below by a constant as $P\to\infty$ as well. Thus, $R_s^E/\log(P) \to 0$ and the proof is complete.	
\end{IEEEproof}

\section{Proof of Lemma \ref{lemma:leakage_power}}

\begin{IEEEproof}
The transmitter has access to $\hat{\bF}$, a quantized version of $\bF$, which is obtained using (\ref{eq:F_design}). The transmitter designs $\bW_{2,Q}$ such that $\hat{\bF}^*\bW_{2,Q}=0$ (recall (\ref{eq:F_constraint})). Letting $\hat{\bM}=\hat{\bF}\bU$, we also have $\hat{\bM}^*\bW_{2,Q} = 0$. Using this, we obtain:
\begin{align*}
	\bG^*\bV^*\bH_d\bW_{2,Q} &= \bG^* \bV^* \bC^* \bF^* \bW_{2,Q} \\
		&= \bB^* \bF \bC \bV (\bV^* \bC^* \bC \bV)^{-1} (\bC \bV)^* \bF^* \bW_{2,Q}
\end{align*}

Consider the square invertible positive definite matrix
\begin{equation} \label{eq:P_def}
	\bP = \bC \bV (\bV^*\bC^* \bC\bV)^{-1} (\bC \bV)^*
\end{equation}
This matrix can be seen as a projection onto the column space of $\bC\bV$. Write the eigendecomposition of $\bP$ as $\bU \bLambda \bU^*$. Then, continuing:
\begin{align*}
	\bG^*\bV^*\bH_d\bW_{2,Q} &= \bB^* \bF \bP \bF^* \bW_{2,Q} \\
			&= \bB^* \bF \bU \bLambda \bU^* \bF^* \bW_{2,Q} \\
			&= \bB^* \bM \bLambda \bM^* \bW_{2,Q} \\
			&= \bB^* \bM \bLambda \bM^* \bW_{2,Q} - \bB^*\hat{\bM} \hat{\bM}^* \bW_{2,Q} \\
			&= \bB^* (\bM \bLambda \bM^* - \hat{\bM} \hat{\bM}^*) \bW_{2,Q}
\end{align*}
The leakage power can then be bounded as:
\begin{align}
	L &= \EE \nn \sqrt{1-\rho} \bG^* \bV^* \bH_d \bW_{2,Q} \bx_{an} \nn_2^2 \nonumber \\
		&\leq (1-\rho) \EE\nn \bx_{an} \nn_2^2 \nn \bG^*\bV^* \bH_d \bW_{2,Q}\nn_F^2  \label{eq:L_ub1} \\
		&= \frac{(1-\rho) P}{n_t-n_r} \nn \bB^* (\bM \bLambda \bM^* - \hat{\bM} \hat{\bM}^*) \bW_{2,Q} \nn_F^2  \nonumber \\
		&\leq \frac{(1-\rho) P}{n_t-n_r} \nn \bB^* \nn_2^2 \nn \bM \bLambda \bM^* - \hat{\bM} \hat{\bM}^*\nn_F^2 \nn \bW_{2,Q} \nn_2^2 \nonumber \\
		&= \frac{(1-\rho) P}{n_t-n_r} \nn \bM \bLambda \bM^* - \hat{\bM} \hat{\bM}^*\nn_F^2 	\label{eq:L_bnd1}
\end{align}
where we used the fact that $\nn \bB^* \nn_2 = \nn \bW_{2,Q} \nn_2 = 1$ since they are truncated unitary matrices.

Since $\bP$ is a projection matrix, it follows its eigenvalues $[\bLambda]_{i,i}$ are either $1$ or $0$. As a result, we have:
\begin{equation} \label{eq:e2}
	\nn \bM \bLambda \bM^*\nn_F^2 \leq \nn \bM\bM^*\nn_F^2 
\end{equation}
and since $\bI-\bLambda=:\bD$ consists of zeros or ones on the diagonal, we also have:
\begin{align}
	\tr(\hat{\bM}^*\bM(\bI-\bLambda)(\hat{\bM}^*\bM)^*) &= \tr(\tilde{\bM}\bD\tilde{\bM}^*) = \tr(\bD \tilde{\bM}^*\tilde{\bM}) \nonumber \\
		&=\sum_{i} [\bD]_{i,i} [\tilde{\bM}^*\tilde{\bM}]_{i,i} \geq 0 \label{eq:e3}
\end{align}
Using (\ref{eq:e2}) and (\ref{eq:e3}), we obtain:
\begin{equation} \label{eq:e4}
	\nn\bM\bLambda\bM^*-\hat{\bM}\hat{\bM}^*\nn_F^2 \leq \nn\bM\bM^*-\hat{\bM}\hat{\bM}^*\nn_F^2
\end{equation}
Using the bound (\ref{eq:e4}) in (\ref{eq:L_bnd1}), we obtain:
\begin{align}
	L &\leq \frac{(1-\rho) P}{n_t-n_r} \nn\bM\bM^*-\hat{\bM}\hat{\bM}^*\nn_F^2 \nonumber \\
		&= \frac{(1-\rho) P}{n_t-n_r} \nn\bF\bF^*-\hat{\bF}\hat{\bF}^*\nn_F^2 \nonumber \\
		&= \frac{2(1-\rho) P}{n_t-n_r} d_c(\bF,\hat{\bF})^2 \label{eq:L_bnd2}
\end{align}
where we used $\bU\bU^*=\bI_{n_r}$ and the definition of the chordal distance.

Assuming a sphere-packing codebook construction, Thm. 5 in \cite{Krishnamachari:2011} yields a bound on the maximum quantization error:
\begin{equation} \label{eq:quantization_bnd}
	\max_{\bF \in \mathcal{G}_{n_t,n_r}} d_c(\bF,\hat{\bF}) \leq \frac{2}{(c2^{N_f})^{1/N}} (1+o(2^{-\frac{N_f}{N}}))
\end{equation}
where $\hat{\bF}$ is obtained using (\ref{eq:F_design}) and $c$ is the coefficient of the ball volume in $\mathcal{G}_{n_t,n_r}$ \footnote{The constant $c$ is given by \cite{Dai:2008}:
\begin{equation*}
	c = \frac{1}{(n_r(n_t-n_r))!} \prod_{i=1}^{n_r} \frac{(n_t-i)!}{(n_r-i)!}
\end{equation*}
}. Using the quantization error bound (\ref{eq:quantization_bnd}) in (\ref{eq:L_bnd2}), we obtain the desired bound in (\ref{eq:L_bnd}).
\end{IEEEproof}

\section{Proof of Corollary \ref{cor:L_bounded}}

\begin{IEEEproof}
	With the choice $N_f=N/2\log_2 P$, Lemma \ref{lemma:leakage_power} implies 
	\begin{align*}
		L &\leq \frac{2(1-\rho)P}{n_t-n_r} \frac{4/c^2}{P} (1+o(P^{-1/2})) \\
			&\stackrel{P\to\infty}{\longrightarrow} \frac{8(1-\rho)/c^2}{n_t-n_r}
	\end{align*}
	The proof is complete.
\end{IEEEproof}

\section{Proof of Lemma \ref{lemma:beta_P}}

\begin{IEEEproof}
We want to show that $\beta(P)$ is positive asymptotically as $P\to\infty$. Define the matrices
\begin{align*}
	\bM_1(P) &= \rho P \bG^* \bV^* \bH_d \bW_1\bW_1^* \bH_d^*\bV\bG \\
	\bM_2(P) &= (1-\rho)\frac{P}{n_t-n_r} \bG^*\bV^*\bH_d\bW_{2,Q}\bW_{2,Q}^*\bH_d^* \bV\bG \\
	\bA(P)   &= \bM_1(P) + \bM_2(P) + \sigma^2 \bG^*\bG \\
	\bGamma(P)  &= \bW_{1,Q}\bW_{1,Q}^* - \bW_1\bW_1^* \\
	\bZ      &= \bG^*\bV^*\bH_d
\end{align*}
and the scalar $\delta=\rho P$. Then, we can write:
\begin{equation*}
	\beta(P)=\log\det(\bA(P) +\delta(P) \bZ\bGamma(P)\bZ^*) - \log\det(\bA(P)).
\end{equation*}
Recall the choice $\bW_{1,Q}=\hat{\bF}$ and $\bW_1=\bF$. Thus, the matrix $\bGamma(P)$ can be rewritten as:
\begin{equation*}
	\bGamma(P) = \hat{\bF}\hat{\bF}^* - \bF\bF^*
\end{equation*}
If $\hat{\bF}\hat{\bF}^* \succeq \bF\bF^*$, then it follows that $\beta(P)\geq 0$ for all $P$, but there is not necessarily true in general. Instead, we show $\beta(P)\geq 0$ for large $P$. Using the perturbation Lemma \ref{lemma:logdet_perturbation} with $\bDelta(P) = \delta(P) \bZ\bGamma(P)\bZ^*$, we obtain
\begin{align}
	\beta(P) &\geq \tr(\bDelta(P) (\bA(P)+\bDelta(P))^{-1}) \log(e) \nonumber \\
		&\geq -\nn\bDelta(P) \nn_F \nn (\bA(P)+\bDelta(P))^{-1}\nn_F \log(e) \label{eq:beta_P_lb}
\end{align}
We conclude the proof by showing that $\nn\bDelta(P) \nn_F \cdot \nn (\bA(P)+\bDelta(P))^{-1}\nn_F=o(1)$ as $P\to\infty$. To this end, first note the bounds:
\begin{align}
	\nn&\bDelta(P)\nn_F \nonumber \\
	  &= \delta(P) \nn\bZ\bGamma(P)\bZ^*\nn_F \nonumber \\
		&\leq \delta(P) \nn\bZ\nn_2^2 \nn\bGamma(P)\nn_F \nonumber \\
		&= \sqrt{2} \rho P \nn\bZ\nn_2^2 d_c(\bF,\hat{\bF}) \nonumber \\
		&\leq \sqrt{2} P \frac{2 \nn\bZ\nn_2^2 2^{-N_f/N}}{c^{1/N}} (1+o(2^{-N_f/N}))\Bigg|_{N_f=\frac{N}{2} \log_2 P}  \label{eq:Delta_bnd} \\
		&= O(\sqrt{P}) 	\quad (\text{as $P\to\infty$}) \label{eq:Delta_zero}
\end{align}
where we used the upper bound (\ref{eq:quantization_bnd}) on the chordal distance in (\ref{eq:Delta_bnd}). Next, consider the positive semidefinite matrix $\bM_2(P)$. Using the bounds in the proof of Lemma \ref{lemma:leakage_power} (see (\ref{eq:L_ub1})), it follows that:
\begin{align}
	\tr(&\bM_2(P)) \nonumber \\
	  &= \frac{(1-\rho)P}{n_t-n_r} \tr(\bG^*\bV^*\bH_d \bW_{2,Q} \bW_{2,Q}^* \bH_d^* \bV \bG) \nonumber \\
		&= \frac{(1-\rho)P}{n_t-n_r} \nn \bG^*\bV^*\bH_d \bW_{2,Q} \nn_F^2 \nonumber \\
		&\leq \frac{2(1-\rho)P}{n_t-n_r} d_c(\bF,\hat{\bF})^2 \nonumber \\
		&\leq \frac{4(1-\rho)P}{c^{2/N}(n_t-n_r)} 2^{-2N_f/N} (1+o(2^{-N_f/N})) \Bigg|_{N_f=\frac{N}{2} \log_2 P} \nonumber \\
		&\leq \frac{4(1-\rho)}{c^{2/N}(n_t-n_r)} (1+o(P^{-1/2})) \nonumber \\
		&= O(1) \quad (\text{as $P\to\infty$}) \label{eq:tr_M2}
\end{align}
On the other hand, the sequence of matrices $\bM_1(P)$ converges to infinity as $P\to\infty$ since $\bM_1(P)=\rho P \bK_{\text{const}}$ for some constant strictly positive definite matrix $\bK_{\text{const}}$ \footnote{In fact, the matrix $\bK_{\text{const}}$ can be shown to be
\begin{equation*}
	\bK_{\text{const}} = \bG^*\bV^*\bC^*\bC\bV\bG = \bB^* \bM \bLambda \bM^* \bB
\end{equation*}
where we used the definition of $\bG$ from (\ref{eq:G_def}), $\bM=\bF\bU$, the eigendecomposition of $\bP$ from (\ref{eq:P_def}) and the QR decomposition $\bH_d^*=\bF\bC$.
}. Thus, we have:
\begin{align}
	\nn& (\bA(P)+\bDelta(P))^{-1} \nn_F \nonumber \\
		&= \nn \left( \rho P \bK_{\text{const}} + \bM_2(P) + \sigma^2\bG^*\bG + \bDelta(P) \right)^{-1} \nn_F \nonumber  \\
		&= P^{-1} \nn \left(\rho \bK_{\text{const}} + \left\{ \frac{\bM_2(P) + \sigma^2 \bG^*\bG + \bDelta(P)}{P} \right\} \right)^{-1} \nn_F 		\label{eq:eAD}
\end{align}
Next, we notice that the term in the brackets above is $O(P^{-1/2})$ since $\bG$ is independent of $P$ and from (\ref{eq:tr_M2}) and (\ref{eq:Delta_zero}):
\begin{align*}
	\tr\left( \frac{\bM_2(P)}{P} \right) &= O(P^{-1}) \\
	\parallel \frac{\bDelta(P)}{P} \parallel_F &= O(P^{-1/2})
\end{align*}
Since the trace of the sequence of positive semidefinite matrices $\{\bM_2(P)\}$ tends to zero, the sequence of the matrices must converge to zero as well at the same rate \cite{HornJohnson:1990}. Substituting these back into (\ref{eq:eAD}), we obtain:
\begin{align}
	&\nn\bDelta\nn_F \nn (\bA(P)+\bDelta(P))^{-1}\nn_F \nonumber \\
		&\quad = O(\sqrt{P} P^{-1} \nn \left( \rho\bK_{\text{const}} + O(P^{-1/2})\bI \right)^{-1} \nn_F) \nonumber \\
		&\quad = O(P^{-1/2}) = o(1) \label{eq:lb_zero}
\end{align}
We thus conclude from (\ref{eq:lb_zero}) and the lower bound (\ref{eq:beta_P_lb}) that choosing $\epsilon(P)=\nn\bDelta(P) \nn_F \nn (\bA(P)+\bDelta(P))^{-1}\nn_F \log(e)$ yields the desired lower bound since $\epsilon(P)$ converges to zero and $\beta(P) \geq -\epsilon(P)$. The proof is complete.
\end{IEEEproof}

\section{Proof of Theorem \ref{thm:quantized_CSI}}

\begin{IEEEproof}
Let $R_{s,G}^P$ denote the achievable rate assuming perfect CSI and $R_{s,G}^Q$ denote the achievable rate under the quantized CSI communication scheme.

The first part of the Theorem will be proven first. With perfect CSI, a similar argument as the one presented in Theorem \ref{thm:perfect_CSI} can be used to show:
\begin{equation} \label{eq:sdof_perfect_CSI}
	\frac{R_{s,G}^P}{\log P} \stackrel{P\to\infty}{\longrightarrow} d_s 
\end{equation}
Of course, the fact that $\bG^*\bG$ is full rank (a.s.) is also used.

Define the secrecy rate difference:
\begin{align}
	&\Delta R_{s,G} := R_{s,G}^P - R_{s,G}^Q \nonumber \\
	&\quad = T_{+}^P - T_{-}^P - T_{+}^Q + T_{-}^Q \label{eq:eR}
\end{align}
where the terms $T_{+}^P, T_{-}^P, T_{+}^Q, T_{-}^Q$ are defined in (\ref{eq:RsP}) and (\ref{eq:RsQ}). We note that the term $-T_{-}^P+T_{-}^Q$ is asymptotically negligible since $-T_{-}^P+T_{-}^Q=o(1)$ as $P\to\infty$. To see this, note:
\begin{align*}
	&T_{-}^P = \log\det\Bigg(\bI + \frac{\rho}{n_r} \bH_e \bW_1\bW_1^* \bH_e^* \\
		&\qquad \times \left(\frac{1-\rho}{n_t-n_r} \bH_e\bW_2\bW_2^*\bH_e^* + \frac{\bar{\sigma}^2}{P} \bI_{n_e} \right)^{-1}  \Bigg) \\
		&\stackrel{P\to\infty}{\longrightarrow} \log\det\Bigg(\bI + \frac{\rho}{1-\rho} \frac{n_t-n_r}{n_r} \bH_e \bW_1\bW_1^* \bH_e^* \\
		&\qquad  \times \left( \bH_e\bW_2\bW_2^*\bH_e^* \right)^{-1}  \Bigg) =: t_\infty
\end{align*}
where we used the condition $n_e\leq n_t-n_r$, implying that $\bH_e\bW_2\bW_2^*\bH_e^*$ is invertible. Using the fact that $\lim_{P\to\infty} N_f = \infty$, it follows that $\bW_{1,Q}\to\bW_1$ and $\bW_{2,Q}\to\bW_2$ as the loss due to quantization becomes asymptotically negligible. Therefore, using a similar technique and taking the limit as $P\to\infty$, it follows that $T_{-}^Q \to t_{\infty}$. Thus, we have $-T_{-}^P+T_{-}^Q \to -t_\infty + t_\infty = 0$. Using this in (\ref{eq:eR}), we obtain:
\begin{equation} \label{eq:RsG_o}
	\Delta R_{s,G} = T_{+}^P - T_{+}^Q + o(1)
\end{equation}
Without loss of generality, let us assume $\bK_{x_s} \sim c P \bI$ for the purposes of analysis since $c_0 P \bI \preceq \bK_{x_s} \preceq c_1 P \bI$. Using the definitions of $T_{+}^P$ and $T_{+}^Q$, after some algebra, we obtain:
\begin{align*}
	&\Delta R_{s,G} \\
	&\sim \log\det\left( \underbrace{\frac{(1-\rho)P}{n_t-n_r} \bG^*\bV^*\bH_d \bW_{2,Q}\bW_{2,Q}^*\bH_d^* \bV \bG}_{\bM_2(P)} + \sigma^2 \bG^*\bG \right) \\
	&\qquad - \log\det(\sigma^2 \bG^*\bG) \\
	&- \Big[ \log\det\Big( \rho P \bG^*\bV^*\bH_d \bW_{1,Q}\bW_{1,Q}^* \bH_d^* \bV\bG \\
	&\quad + (1-\rho)\frac{P}{n_t-n_r} \bG^*\bV^*\bH_d\bW_{2,Q}\bW_{2,Q}^*\bH_d^* \bV\bG + \sigma^2 \bG^*\bG \Big) \\
	&\quad - \log\det\left( \rho P \bG^*\bV^*\bH_d \bW_{1,Q}\bW_{1,Q}^* \bH_d^* \bV\bG + \sigma^2 \bG^*\bG  \right)  \Big] \\
	&\quad + o(1)
\end{align*}
The term in the brackets is exactly the remainder term $\beta(P)$ defined in Lemma \ref{lemma:beta_P}. Using the result of Lemma \ref{lemma:beta_P}, there exists a sequence $\epsilon(P)$ converging to zero such that:
\begin{align}
	&\Delta R_{s,G} \sim \log\det\left( \bM_2(P) + \sigma^2 \bG^*\bG \right) \nonumber \\
		&\qquad - \log\det(\sigma^2 \bG^*\bG) - \beta(P) + o(1) \nonumber \\
		&\leq d_s \log(\nn\bM_2(P)+\sigma^2\bG^*\bG\nn_2) \nonumber \\
		&\qquad - d_s \log(\sigma^2 \lambda_{min}(\bG^*\bG)) + \epsilon(P) + o(1) \nonumber \\
		&\leq d_s \log\left( \underbrace{\frac{\lambda_{max}(\bG^*\bG)}{\lambda_{min}(\bG^*\bG)}}_{\kappa(\bG^*\bG)} +  \frac{\nn \bM_2(P) \nn_2}{\sigma^2 \lambda_{min}(\bG^*\bG)} \right) + \epsilon(P)+o(1) \nonumber \\
		&\leq d_s \log\left( \kappa(\bG^*\bG) + \frac{(1-\rho) P \nn \bG^*\bV^*\bH_d \bW_{2,Q}\nn_F^2}{\sigma^2 \lambda_{min}(\bG^*\bG)} \right) \nonumber \\
		&\qquad + \epsilon(P) + o(1) \label{eq:R_sG_bnd}
\end{align}
The term given by $P \nn \bG^*\bV^*\bH_d \bW_{2,Q}\nn_F^2$ is $O(1)$ as $P\to\infty$, as the bounds in (\ref{eq:tr_M2}) show. Thus, dividing both sides of (\ref{eq:R_sG_bnd}) by $\log P$ and taking the limit as $P\to\infty$, we obtain:
\begin{align*}
	d_s^Q &:= \lim_{P\to\infty} \frac{R_{s,G}^Q}{\log P} \\
		&\geq \lim_{P\to\infty} \frac{R_{s,G}^P}{\log P} - \lim_{P\to\infty} \frac{d_s \log(\kappa(\bG^*\bG) + O(1))+\epsilon(P)+o(1)}{\log P} \\ 
		&= d_s - 0 = d_s
\end{align*}
where we used (\ref{eq:sdof_perfect_CSI}). Since the rate of the quantized CSI scheme is less than the rate of the perfect CSI scheme, it follows that $d_s^Q\leq d_s$. Thus, we conclude that $d_s^Q=d_s$.

The second part of the Theorem now easily follows. Note that from (\ref{eq:RsG_o}), we have:
\begin{align}
	\delta_{\text{GAP}} &= \lim_{P\to\infty} \{R_{s,G}^P - R_{s,G}^Q\} \nonumber \\
		&= \lim_{P\to\infty} \{ T_+^P - T_+^Q \} \nonumber \\
		&\leq \lim_{P\to\infty} \log\det\left( \bI + \frac{(1-\rho)}{\sigma^2}\bU(P) (\bG^*\bG)^{-1} \right) \label{eq:delta_zero}
\end{align}
where
\begin{equation*}
	\bU(P) := P \bG^* \bV^* \bH_d \bW_{2,Q} \bW_{2,Q}^* \bH_d^* \bV \bG
\end{equation*}
is a positive semidefinite matrix depending on $P$. From the development in (\ref{eq:tr_M2}), using the scaling in (\ref{eq:epsilon_Nf}), we obtain:
\begin{align*}
	\tr(\bU(P)) &= P \nn \bG^* \bV^* \bH_d \bW_{2,Q} \nn_F^2 \\
		&\leq 2 P d_c(\bF,\hat{\bF})^2 \\
		&\leq \frac{8 P}{c^{2/N}} 2^{-2N_f/N} (1+o(2^{-\frac{N_f}{N}})) \Bigg|_{N_f=\frac{(1+\epsilon)N}{2} \log_2 P} \\
		&= O(P^{-\epsilon})
\end{align*}
As a result, $\tr(\bU(P))=o(1)$, implying $\bU(P) \to \mathbf{0}$ as $P\to\infty$ \cite{HornJohnson:1990}. Using this result in (\ref{eq:delta_zero}), we have $\delta_{\text{GAP}}\leq 0$. The final result following by noting that $\delta_{\text{GAP}}$ is always nonnegative. The proof is complete.
\end{IEEEproof}


%


\bibliographystyle{IEEEtran}
\bibliography{refs}

\end{document}